\documentclass[aps,prl,twocolumn,superscriptaddress,floatfix,10pt]{revtex4-1}
\usepackage{graphicx}
\usepackage{dcolumn}
\usepackage{bm}
\usepackage{amsfonts}
\usepackage{amssymb}
\usepackage{amsmath}

\usepackage{bbold}

\newcommand{\oeq}{\begin{equation}}
\newcommand{\ceq}{\end{equation}}
\newcommand{\oeqn}{\begin{eqnarray}}
\newcommand{\ceqn}{\end{eqnarray}}

\renewcommand{\]}{\right]}

\newcommand{\hb}{\hbar}

\begin{document}

\title{Interplay between quantum shells and orientation in quasi-fission}

\author{A. Wakhle}
\affiliation{Department of Nuclear Physics, Research School of Physics and Engineering, Australian National University, Canberra, Australian Capital Territory 0200, Australia}

\author{C. Simenel}
\affiliation{Department of Nuclear Physics, Research School of Physics and Engineering, Australian National University, Canberra, Australian Capital Territory 0200, Australia}

\author{D. J. Hinde}
\affiliation{Department of Nuclear Physics, Research School of Physics and Engineering, Australian National University, Canberra, Australian Capital Territory 0200, Australia}

\author{M. Dasgupta}
\affiliation{Department of Nuclear Physics, Research School of Physics and Engineering, Australian National University, Canberra, Australian Capital Territory 0200, Australia}

\author{M. Evers}
\affiliation{Department of Nuclear Physics, Research School of Physics and Engineering, Australian National University, Canberra, Australian Capital Territory 0200, Australia}

\author{D. H. Luong}
\affiliation{Department of Nuclear Physics, Research School of Physics and Engineering, Australian National University, Canberra, Australian Capital Territory 0200, Australia}

\author{R. du Rietz}
\affiliation{Department of Nuclear Physics, Research School of Physics and Engineering, Australian National University, Canberra, Australian Capital Territory 0200, Australia}

\author{E. Williams}
\affiliation{Department of Nuclear Physics, Research School of Physics and Engineering, Australian National University, Canberra, Australian Capital Territory 0200, Australia}

\date{\today}

\begin{abstract}
The quasi-fission mechanism hinders fusion in heavy systems through breakup within zeptoseconds into two fragments with partial mass equilibration. Its dependence on the structure of both the collision partners and the final fragments is a key question. Our original approach is to combine an experimental measurement of the fragments' mass-angle correlations in $^{40}$Ca$+^{238}$U with microscopic quantum calculations.
We demonstrate an unexpected interplay between the orientation of the prolate deformed  
 $^{238}$U with quantum shell effects in the fragments. In particular, calculations show that only collisions with the tip of $^{238}$U produce quasi-fission fragments in the magic $Z=82$ region, whilst collisions with the side are the only one which may result in fusion. 
\end{abstract}
\maketitle


In the late 70's, Heusch and collaborators measured fission characteristics in heavy-ion collisions which could not be reconciled with the statistical decay of a compound nucleus \cite{heu78}.
Later, the angular anisotropy of the fission fragments was found to be much larger than that predicted by the statistical model in some reactions \cite{bac81,boc82}, which was taken as a clear signature for an out-of-equilibrium process.

The origin of these characteristics is understood to be a process known as quasi-fission. 
Here 
the dinuclear system fissions before reaching the stage of an equilibrated compound nucleus \cite{boc82}. 
Quasi-fission thus results in fusion hindrance in reactions forming heavy nuclei \cite{sah84,gag84,sch91}.
In fact, this is by far the dominant mechanism suppressing the formation of super-heavy elements. 
The understanding of this process is thus crucial in order to optimise the formation of new heavy and superheavy nuclei. 

Since the discovery of quasi-fission, important progress has been made thanks to extensive experimental studies \cite{tok85,she87,hin92,hin95,hin96,itk04,kny07,hin08,nis08,koz10,rie11,itk11,lin12,nis12,sim12a,rie13,wil13,koz14}. 
Correlations between the mass and the angles of the fragments show that quasi-fission often takes place before a full rotation of the di-nuclear system, that is, with typical contact times between the fragments of 5 to 10~zs \cite{tok85,she87,rie11}.
The characteristics of the entrance channel - in particular, the deformation \cite{hin95,hin96,kny07,hin08,nis08} and shell structure \cite{sim12a} of the collision partners as well as the fissility of the system \cite{lin12,rie13}  and its energy \cite{nis08,nis12} - were shown to play an important role. 
Shell effects could also favor the production of fragments in the vicinity of magic nuclei \cite{itk04,nis08,mor08,fre12,koz10,koz14}.

The complex interplay of all these variables that have been identified by experiments 
dictates quasi-fission characteristics and probability, and hence the suppression of fusion. 
To understand the  dynamics at play, in particular the inter-dependency of these variables, it is necessary to perform theoretical calculations.
Classical dynamical models have  been  developed where the  system is described as a viscous fluid evolving through a family of parametrised shapes \cite{dia01,ada03,zag05,ari09,min10,ari12}.
Despite their ability to reproduce some experimental observables, these approaches require parameters, such as the viscosity, which must be provided externally. 
One possibility is to extract these parameters directly from microscopic approaches \cite{was09a,wen13}, for which the only parameters are those describing the interaction between nucleons. 
Here, we take another approach using a quantum microscopic model to  directly investigate  quasi-fission dynamics. 
Although computationally more demanding, microscopic calculations have the advantage of not constraining the shape of the system during its evolution. 
 The quantum aspect of the model is also crucial to investigate the role of shell effects in the dynamics. 
The theoretical analysis of the experimental data presented in this letter is performed with modern microscopic calculations based on the time-dependent Hartree-Fock (TDHF) theory. The latter has recently been successful in describing dynamical processes such as vibration \cite{sim03,mar05,nak05,ave13}, fusion \cite{sim01,uma06a,was08,guo12,sim13b,sim13c,uma14}, transfer reactions \cite{sim10b,sca13,sek13}, deep-inelastic collisions \cite{sim11}, and quasi-fission in actinide collisions \cite{gol09,ked10} (see Ref.~\cite{sim12b} for a review), as well as the dynamics of fission fragments \cite{sim14}.

Mass and angle distributions (MAD) of fission fragments formed in $^{40}$Ca+$^{238}$U collisions have been measured and calculated at different energies to investigate the role of quantum shell effects on the final characteristics of the fragments.
In particular, we answer key questions, such as the interplay between the orientation of a deformed collision partner and the quantum shells affecting the outcome of the quasi-fission reaction. 
 The findings are relevant to understanding the dynamics for forming the next superheavy elements since all the available targets that are planned to be used are deformed.

Pulsed beams of $^{40}$Ca were produced using a 14UD electrostatic accelerator followed by a LINAC post-accelerator at the Australian National University.
Isotopically enriched targets of $^{238}$UF$_{4}$ (250~$\mu$g/cm$^2$), evaporated onto $\sim$15~$\mu$g/cm$^2$ $^{nat}$C backings, were mounted
on a target ladder whose normal was at 60$\,^{\circ}$ to the beam.
Binary reaction products were detected in coincidence using two 28$\times$36~cm$^2$ position-sensitive multiwire proportional
counters on opposite sides of the beam, covering laboratory scattering angles of
$5\,^{\circ}<\theta<80\,^{\circ}$ and $50\,^{\circ}<\theta<125\,^{\circ}$.
The measured positions and times-of-flight allowed direct reconstruction of the fragment velocities~\cite{hin96}. 
The latter were converted into mass ratio $M_R=m_1/(m_1+m_2)$ and  center-of-mass (c.m.) scattering angle $\theta_{c.m.}$  for events where only two primary fragments with masses $m_1$ and $m_2$ are formed.
The selection of full momentum transfer (binary) events is described in details in Ref.~\cite{rie13}.
Since both fragments are detected, the MAD is populated twice~\cite{tho08}, at ($M_R, \theta_{c.m.})$ and $(1-M_R, \pi-\theta_{c.m.}$).

\begin{figure}[!htb]
\includegraphics*[width=8.5cm]{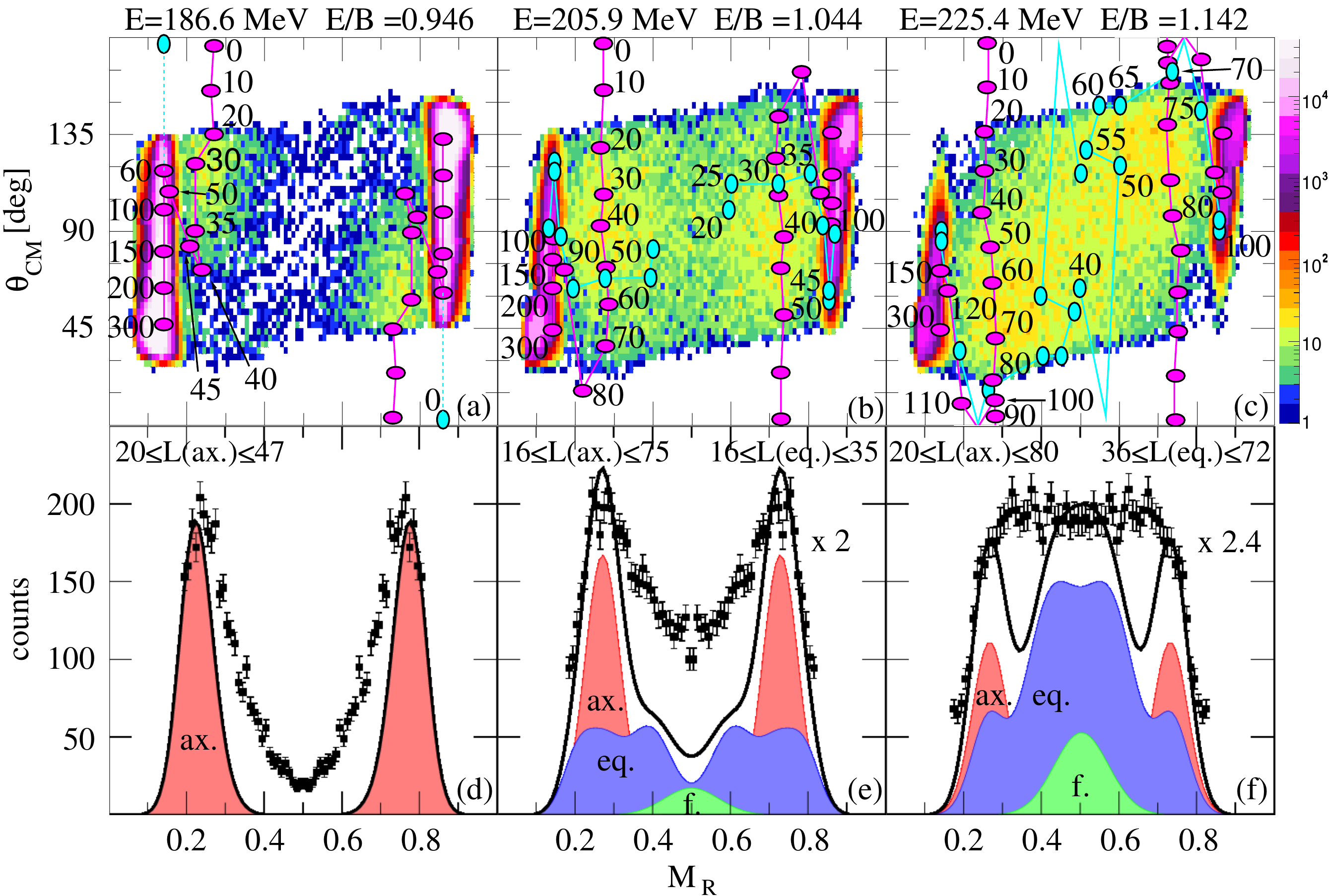}
\caption{(Color online) (a-c) Measured mass-angle distributions at center of mass energies $E$ near the capture barrier
$B$  \cite{swi05}.
The logarithmic colour scale is shown in the top-right. 
The horizontal and vertical ellipsoids show the TDHF results for the axial and equatorial configurations, respectively. 
The values of $L$ are indicated near the points associated to the light (heavy) fragments for the axial (equatorial) orientation.
(d-f) Projected mass-ratio $M_R$ in the range $0.2<M_R<0.8$. 
The scale factor in panels (e) and (f) multiplies the counts scale on the left. 
The mass ratio distributions estimated from the TDHF results are shown with shaded areas for quasi-fission in the axial (ax.) and equatorial (eq.) configurations, and for fusion-fission (f.) events. 
The ranges of $L$ used to calculate the mass ratio distributions are given in panels (d-f) and correspond to events falling into the angular acceptance of the detector.
The sum of these distributions is shown (solid line). 
}
\label{fig:MAD}
\end{figure}

The MAD measured at three energies are shown in Fig.~\ref{fig:MAD}(a-c). 
The azimuthal coincidence coverage of the detector system was 90$\,^{\circ}$ for all~$\theta_{c.m.}$;
thus, the number of events in each MAD bin is proportional to the angular differential cross section d$\sigma$/d$\theta_{c.m.}$.
No events were detected at the most forward and backward angles due to detector angular acceptance.
The intense bands at extreme $M_R$ values correspond to (quasi-)elastic scattering.
Events associated with quasi-fission and fusion-fission are located between these two bands.

Each MAD shows very mass-asymmetric groups of fission events, with the light fragment in the range $M_R\simeq0.2-0.3$ and a corresponding heavy fragment. These groups move toward forward and backward angles, respectively, with increasing energy. 
This correlation of mass with angle is a clear signature for quasi-fission events. 
In addition, the fraction of events around $M_R=0.5$ (indicating  mass equilibration of the fragments) increases with energy. 
These mass symmetric events are compatible with long life time quasi-fission (i.e., with contact times  greater than half a rotation), or fusion followed by statistical fission. 

The projections of the MADs onto the $M_R$ axis are shown with filled squares in Figs.~\ref{fig:MAD}(d-e). 
Peaks at $M_R\simeq0.25$ and 0.75, associated with asymmetric quasi-fission, are clearly visible at the two lowest energies. 
At the highest energy, less asymmetric events dominate and a wide plateau is observed between $M_R\simeq0.3$ and 0.7, in excellent agreement with mass distributions presented in~\cite{nis12}.

To interpret these observations, TDHF calculations were performed at the same energies using the \textsc{tdhf3d} code \cite{kim97}. 
The system is described by an anti-symmetrised independent particle state at all time to ensure an exact treatment of the Pauli principle, which is crucial at low energies. 
The TDHF equation  is $i\hb\frac{d}{dt}\rho=\[h[\rho],\rho\]$, where $\rho$ is the one-body density matrix \cite{dir30}.
The Hartree-Fock (HF) Hamiltonian $h[\rho]$ is obtained from a Skyrme energy density functional \cite{sky56}.
Unlike early calculations of similar reactions which used simplified interactions \cite{sto82}, the SLy4$d$ parametrisation \cite{kim97} used here includes a spin-orbit interaction \cite{cha98}.
The latter is crucial to reproduce the one-body dissipative mechanisms \cite{uma86} which strongly affect low-energy dynamics, as well as magic numbers in heavy nuclei. 
More details can be found in Ref.~\cite{sim12b}.

\begin{figure}[!htb]
\includegraphics*[width=8cm]{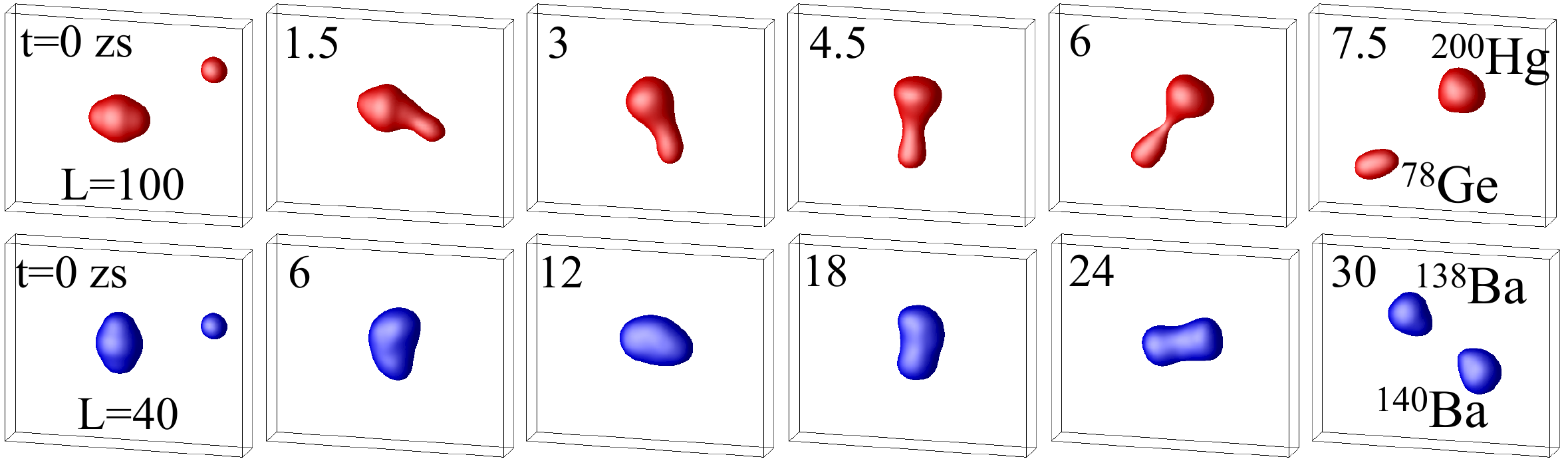}
\caption{(Color online) Time evolution of the density in $^{40}$Ca+$^{238}$U collisions at a center of mass energy $E=225.4$~MeV. 
The isodensity at half the saturation density $\rho_0/2=0.08$~fm$^{-3}$ is plotted every 1.5~zs for the axial orientation at $L=100$ (top) and every 6~zs for the equatorial orientation at $L=40$ (bottom).}
\label{fig:snapshots}
\end{figure}

Examples of shape evolution for $^{40}$Ca$+^{238}$U at $E=225.4$~MeV 
are shown in Fig.~\ref{fig:snapshots} for two extreme orientations of the prolate $^{238}$U nucleus.
In the collision with the tip of $^{238}$U (axial orientation) at an angular momentum quantum number $L=100$ (upper panels), 
the  system undergoes half a rotation in $\sim5$~zs before it separates into primary fragments with smaller mass asymmetry. 
This relatively short contact time and the partial mass equilibration are clear signatures of a quasi-fission process. 
The collision with the side of $^{238}$U (equatorial orientation) at $L=40$ (lower panels) exhibits a much longer contact time (about 30~zs) and a full rotation before separation, which is also compatible with a quasi-fission process.
However, this time is long enough to induce a full mass equilibration, leading to symmetric mass-split.
The total kinetic energy of the final fragments is predicted to be $\sim243$~MeV, 
in excellent agreement with experimental data for $M_R\simeq0.5$ \cite{nis12}. 

Systematic calculations have been performed by varying $L$ with steps $\Delta L=10$ or 5.
See Supplemental Material at [URL will be inserted by publisher] for movies of the density evolution associated with these calculations.
The results are reported on the MADs in Fig.~\ref{fig:MAD}.
Axial (equatorial) orientations are represented by purple horizontal (blue vertical) ellipsoids. 
At the lowest energy [Fig.~\ref{fig:MAD}(a)], axial collisions lead to quasi-fission up to $L\sim45$, and above that to quasi-elastic scattering.
Equatorial orientations  contribute only to the elastic peak. 
The  light quasi-fission fragments are essentially located at backward angles with mass ratio in the range $M_R\sim0.22-0.27$, in good agreement with the data falling into the angular acceptance of the detector. 
Similar conclusions can be drawn at  higher energies [Figs~\ref{fig:MAD}(b) and~(c)] for the axial collisions, with the angles of the light quasi-fission fragments going toward more forward angles with increasing energy. 
Fusion does not occur for the axial orientation.
In contrast, equatorial collisions form a compact system which has not decayed into fission fragments at the end of the calculations for $L\le10$ at $E=205.9$~MeV and $L\le30$ at $E=225.4$~MeV. 
These events, associated with long life time quasi-fission or with fusion-fission, are expected to produce fragments with isotropic angular distributions. They are called "fusion-fission" hereafter. 
Quasi-fission events are obtained at larger angular momenta and include more symmetric events than found in axial collisions.
These comparisons between TDHF predictions and data are meaningful only when fission of the heavy fragment is negligible.
Only fragments significantly heavier than Pb could fission, affecting the $M_R\ge0.76$ and $M_R\le0.24$ regions. 
However, the calculations predict negligible yields of quasi-fission fragments in this $M_R$ region.

\begin{figure}[!htb]
\includegraphics*[width=7cm]{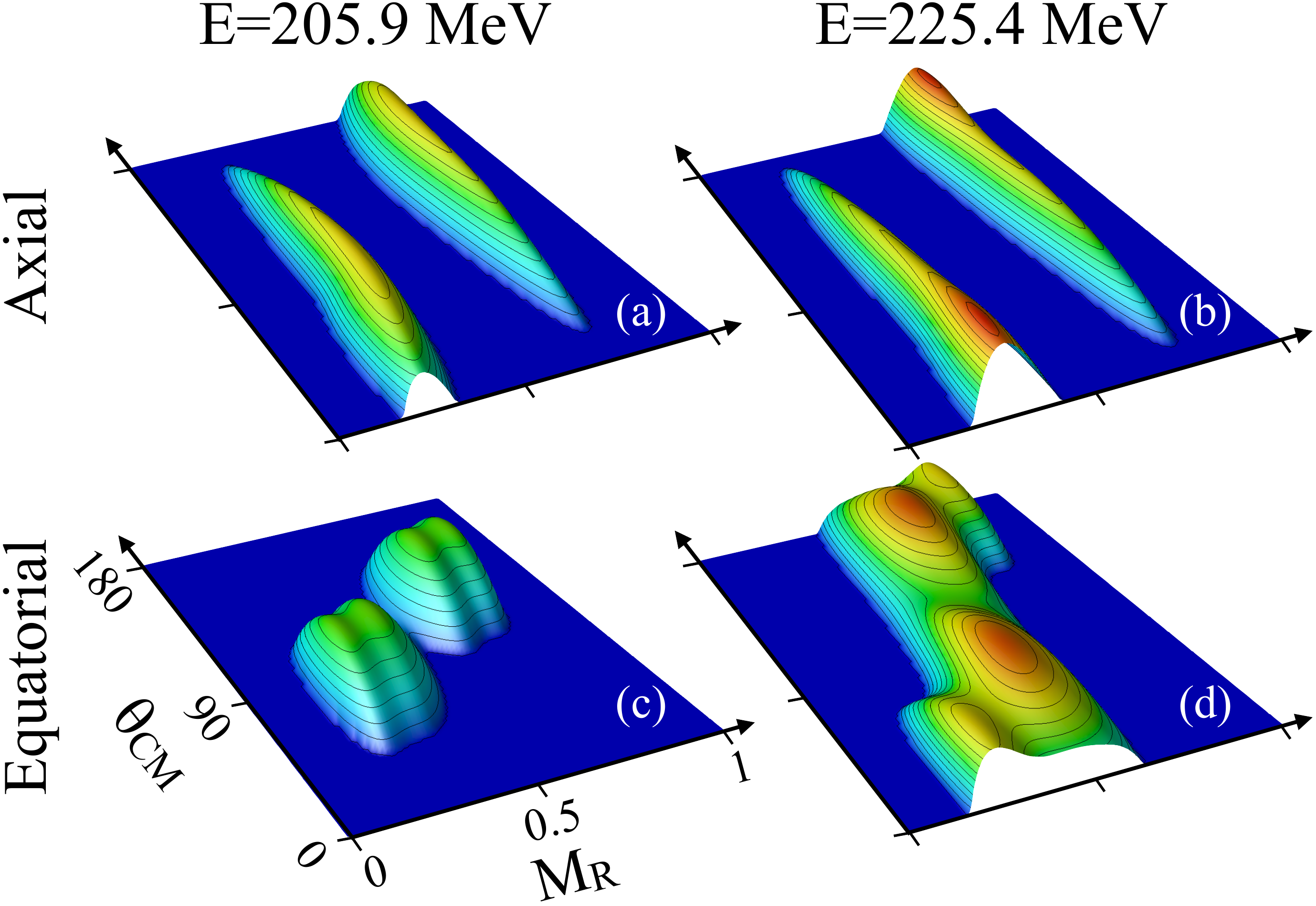}
\caption{(Color online) Mass-angle distributions of quasi fission fragments from TDHF calculations.
The common vertical linear scale corresponds to the angular differential cross section in arbitrary units.}
\label{fig:madsim}
\end{figure}

In order to estimate the overall significance of the orientation of the $^{238}$U nucleus, a representation of the MAD of quasi-fission fragments  is given in Figs.~\ref{fig:madsim}(a-d) at the two highest energies.
Since TDHF calculations underestimate the fluctuation of the fragment mass distributions in damped collisions \cite{das79}, 
Gaussian distributions centered around each $M_R$ and $\theta_{c.m.}$ obtained from TDHF and weighted by $2L+1$ are assumed, with a standard deviation in mass ratio varying linearly with $M_R$ from 0.025 at the initial mass split to 0.07 at symmetry \cite{rie11}, and a standard deviation in angle of $20^\circ$.
The relative weight between the two orientations is determined by assuming that axial collisions are obtained when the angle between the beam axis and the target nucleus symmetry axis is smaller than $35^\circ$ \cite{hin96}. 
The projections on the mass-ratio axis (for the angular acceptance of the detectors) are shown in Figs.~\ref{fig:MAD}(d-f) by shaded areas, 
together with fusion-fission events which are assumed to produce symmetric fragments with isotropic angular distributions. 
The resulting total distributions are normalised to the most mass-asymmetric experimental events. 
Note that a quantitative reproduction of the experimental MAD is beyond the scope of this work, as it 
would require extensive calculations at intermediate orientations. 
It is observed that the axial orientation is mostly responsible for asymmetric quasi-fission at all energies. 
The more symmetric events are  populated by the equatorial collisions which are dominant at the highest energy, giving rise to the observed plateau in Fig.~\ref{fig:MAD}(f).

To test the validity of this approach, we  computed the ratio of the fusion to capture 
cross-sections $\sigma_f/\sigma_c=0.09\pm0.07$ at 205.9~MeV and $0.16\pm0.06$ at 225.4~MeV 
(no fusion is observed at the lowest energy). The uncertainty is 
due to the angular momentum mesh. These results are in good 
agreement with Ref.~\cite{she87} where a ratio $0.11\sim0.06$ was obtained in this energy range. 
The TDHF capture cross-sections at these energies agree with those of Ref.~\cite{she87} 
within the $\sim20\%$ experimental error bars.

\begin{figure}[!htb]
\includegraphics*[width=6cm]{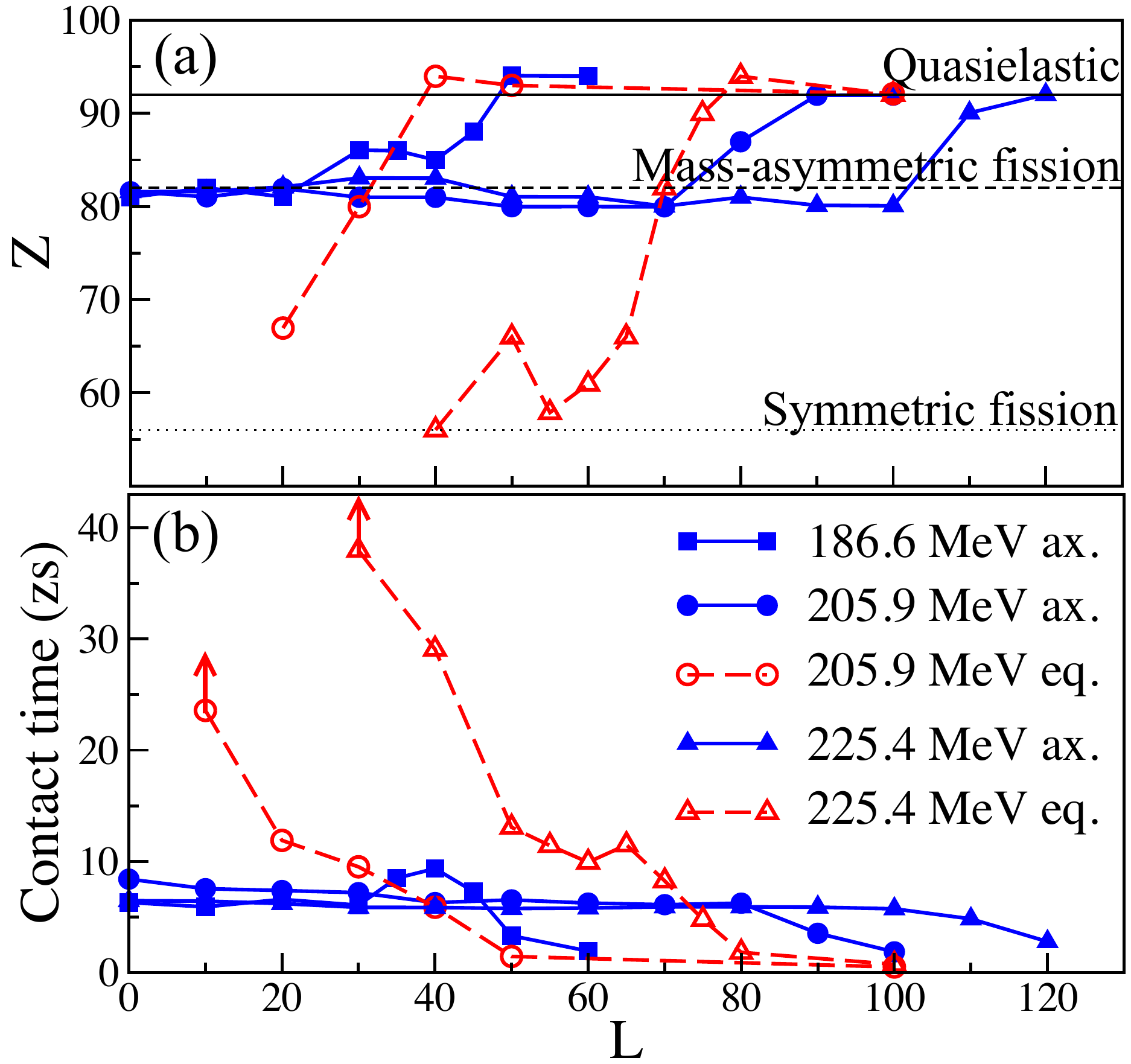}
\caption{ (Color online) (a) Mean number of protons $Z$ in the outgoing heavy fragment as a function of the angular momentum quantum number $L$. Axial and equatorial orientations are represented by thick solid and thick dashed lines, respectively. The horizontal thin solid, thin dashed, and thin dotted lines show the values of $Z$ of the initial $^{238}$U nucleus (quasi-elastic scattering), of the doubly magic $^{208}$Pb nucleus (mass-asymmetric fission), and of the $^{139}$Ba fragment formed in mass-symmetric fission, respectively. (b) Contact time between the fragments defined as the time during which the density in the neck exceeds half the saturation density $\rho_0/2=0.08$~fm$^{-3}$. The arrows indicate lower limits of the contact time.}
\label{fig:ZN}
\end{figure}

We now investigate the role of shell effects in the formation of the fragments. 
The extra binding energy from shell effects is expected to favor the formation of fragments with  magic numbers.
This is supported by the mass-asymmetric quasi-fission observed in the experimental MAD which has its heavy fragment in the $^{208}$Pb region. 
The proton numbers of the heavy fragment calculated in TDHF are plotted as a function $L$ in Fig.~\ref{fig:ZN}(a).
Quasi-fission in axial collisions (filled symbols) always forms a  fragment close to the magic proton number $Z=82$. 
We observed a similar behaviour for neutrons with $N\sim122$, indicating that the influence of the magic number $N=126$ is not as strong as $Z=82$.
These observations do not depend on beam energy. 
However, for equatorial collisions these magic numbers have no visible effects on the outcomes.
One possible reason is that, unlike the axial orientation, equatorial collisions form systems which are more compact than two touching $^{208}$Pb and $^{70}$Zn fragments. 
As shown in Fig.~\ref{fig:ZN}(b), the difference in nucleon transfer between the orientations translates  into  different  quasi-fission times. 
Indeed, the quasi-fission times are smaller than 10~zs and are almost independent of $E$ and $L$ for the axial orientation, while quasi-fission times over 30~zs (see  also Fig.~\ref{fig:snapshots}) are observed for the equatorial orientation. 


Experimental mass-angle distributions of quasi-fission fragments formed in $^{40}$Ca$+^{238}$U 
have been measured and interpreted using TDHF calculations. 
The angular focussing of fragments with large mass asymmetry indicates that they are produced by quasi-fission. 
This asymmetric quasi-fission is related to shell effects in the $Z=82$ region and occurs in collisions with the tip of the $^{238}$U, leading to short quasi-fission times. 
No quantum shell effects were observed in collisions with the side of $^{238}$U. 
Long contact times compatible with fusion are found only for this orientation. 
This first evidence for the orientation dependence of shell effects in quasi-fission requires further investigation. 

Discussions with A. S. Umar and V. E. Oberacker are acknowledged. 
This work has been supported by the
Australian Research Council Grants No. FT120100760, FL110100098 and DP1094947.
The calculations have been performed on the NCI National Facility in Canberra, Australia, which is supported
by the Australian Commonwealth Government.

\bibliography{biblio}

\end{document}